# Evidence for an Oxygen Diffusion Model for the Electric Pulse Induced Resistance Change Effect in Oxides

Y.B. Nian, J. Strozier, N.J. Wu, X. Chen, A. Ignatiev

*Texas Center for Advanced Materials, University of Houston, Houston, TX, 77204*

Abstract: Electric pulse induced resistance (EPIR) switching hysteresis loops for $Pr_{0.7}Ca_{0.3}MnO_3$ (PCMO) perovskite oxide films were found to exhibit an additional sharp "shuttle peak" around the negative pulse maximum for films deposited in an oxygen deficient ambient. The device resistance hysteresis loop consists of stable high resistance and low resistance states, and transition regions between them. The resistance relaxation of the "shuttle peak" and its temperature behavior as well as the resistance relaxation in the transition regions were studied, and indicate that the resistance switching relates to oxygen diffusion with activation energy about 0.4eV. An oxygen diffusion model with the oxygen ions (vacancies) as the active agent is proposed for the non-volatile resistance switching effect in PCMO.

The electric-pulse induced resistance (EPIR) change effect [1] observed in perovskite oxide ($Re_{1-x}A_xMnO_3$, Re=rare-earth ions, $A$=alkaline ions) films has attracted recent extensive interest [2-6]. A perovskite oxide thin film (e.g. $Pr_{0.7}Ca_{0.3}MnO_3$) sandwiched between two conductive electrodes exhibits resistance switching when a short (10-100ns) voltage pulse of several volts is applied across the electrodes. In general the resistance switches down when the current goes into the sample, and switches up when the current comes out of the sample. The resistance switching time is typically $t_s \leq 10^{-7}$ s at a switching voltage of the order of volts, and persists at the high or low resistance state for $t_p > 10^8$ s.

This is the basis for an EPIR memory device, which has great potential for high-speed non-volatile memory. However, the physics of this class of resistance random access memory (RRAM) structures is still under discussion. Switching properties have been associated with properties of the bulk oxide [1], with the bulk near the metal electrode interface [7], with the contact area at the metal/oxide interface [4], and also speculations of a current driven phase separation in the surface layer of the manganite [5]. We report here observations of resistance relaxation with time in the voltage switching region of the EPIR device, and correlate these experiments to a proposed switching model involving oxygen vacancies.

Conduction in PCMO at room temperature is thought to occur by hopping of carriers consisting of small polarons along —Mn—O—Mn—O— chains [8-10] and/or ferromagnetic droplets in a paramagnet insulating matrix [11]. In either case, the process is thermally activated with an activation energy of about 100 – 200 mV where the character of the conduction is thought to be holes [8-10]. We assume that the absence of oxygen atoms (or the presence of oxygen vacancies) in a chain will disrupt the conduction process in that chain by acting as strong scatterers for the 'hole' polarons. An oxygen vacancy might very well trap one or even two electrons forming F or F' like centers. However, the electronic states of these trapped electrons would be more s-like in character and hence less p-like than the atomic orbitals of the oxygen that the vacancies replace. Hence their overlap with the Mn d states would be considerably less, decreasing and possibly eliminating polaron conduction along that chain, hence increasing resistivity. Indeed, the resistivity of $ReAMnO_{3-x}$, such as $LaBaMnO_3$, increased more than four orders of magnitude at room temperature when reduction in the oxygen content increased to ~5% without decomposition [12]. This is also in agreement with the resistance data presented here that films grown in an oxygen-free atmosphere usually have higher resistance than those grown in oxygen.

Further, it is often the case that the surface region of an oxide is oxygen deficient compared to the bulk, containing perhaps 5% to 10% oxygen vacancies in the surface region [13]. Electrodes evaporated or sputtered onto an oxide sample will be in contact with this oxygen deficient surface region, and hence the sample should show high resistivity. This is generally seen in the PCMO device where measured resistance between electrodes deposited on PCMO is much larger than what one would expect from the bulk resistivity of PCMO ($10^{-1}$ to $10^{-2}$ ohm-cm for bulk [14] vs. $10^3$ to $10^4$ ohm-cm for thin film [1,3]). This oxygen deficiency and accompanying vacancies at the electrode interface regions, form the basis for the EPIR switching model to be presented here. We present here data on PCMO thin films with oxygen deficiency, which exhibit resistance relaxation in the EPIR hysteretic switching loop. Through the analysis of the data, we calculate the activation energy and conclude that the oxygen vacancies and their motion play an important role in the EPIR phenomenon.

$Pr_{0.7}Ca_{0.3}MnO_3$ films were grown on bottom-electrode Pt film on $TiN/SiO_2/Si$ substrates by RF

sputtering at 500°C under a 60mTorr Ar only atmosphere (oxygen deficient). 300 µm diameter Ag top-electrode pads were deposited on the PCMO films using DC sputtering for subsequent resistivity measurements. Energy dispersive x-ray spectroscopy (EDS) showed good PCMO film composition and atomic force microscopy (AFM) defined a surface morphology with roughness <10nm. Scanning electron microscopy (SEM) cross-section studies showed PCMO films thickness of ~500nm. Resistance switching of the sample was investigated by applying electrical pulses up to ~5V across the sample electrodes, and the resistance was measured by applying a 1 µA dc current and reading the corresponding voltages cross the film after each applied electric pulse.

Resistance hysteresis switching loops (HSL) of PCMO were measured by applying a sequence of pulses with increasing magnitudes under a constant voltage step (Fig 1.a). Increasing negative pulses were applied when the sample was in its low resistance state until the sample reached the high resistance state (defining the maximum negative pulse voltage). The pulse magnitudes were then reduced until they changed polarity to positive after which the positive pulses were increased in magnitude until the sample switched to the low resistance state (defining the maximum positive pulse voltage). In Fig. 1, the high resistance state identified as region HSL-I and the low resistance state identified as region HSL-III are the EPIR bi-states, and the regions HSL-II and HLS-IV are the resistance transition regions of the EPIR device. For samples grown in an oxygen-free atmosphere, we found that the hysteresis loop shows an unexpected rise to very high resistance values (above the initial high resistance state) under negative pulsing, and then decays to the high resistant state as the negative pulsing voltage is decreased thus producing the 'shuttle tail' peak (the regions HSL-V and HSL-VI in Fig. 1b). This 'peak' is an additional feature to the hysteresis loop not seen for PCMO samples grown in an oxygen environment (Fig. 1a), and we have carefully looked at the dynamic response of the hysteresis curves of R vs. pulse voltage for the Fig. 1b loop in the 'tail' regions HSL-V and HSL-VI, and the transition regions HSL-II and HSL-IV in order to understand the origin of the 'shuttle tail' peak. The resistance as a function of time was measured in the given regions after a voltage pulse was applied, without any additional switching pulses being applied. It was found that in the 'tail' and transition regions (the regions where the resistance was switching) the resistance relaxed back towards the value before the switching pulse was applied (Figs. 2 and 3), i.e., the resistance relaxation is always opposite to the resistance change resulting from the switching pulse. The time constant for this relaxation, $t_r$, is of the order of minutes, which is much larger (by a factor of about $10^9$) than the time constant for switching, $t_s$ (typically about $10^{-7}$ sec for a switching voltage of the order of volts); and much smaller (by a factor of about $10^5$) than the time constant associated with the persistence of the resistance state, $t_p$ ($10^8$ sec or greater). Thus we find there are three essential time constants associated with the EPIR effect: $t_s = 10^{-7}$ s, $t_r = 10^2$ s. and $t_p = 10^8$ s. What is interesting is that the new relaxation process reported here controlled by $t_r$ takes place only in the 'tail' and transition regions, which are the regions where the switching process also occurs. The stable states (the state at high R ($R_H$), the state at low R ($R_L$) and those stable resistance states in between showed no relaxation except where they border the transition region. These states are controlled by $t_p$.

Temperature dependence of the resistance decay was specifically measured in the 'tail' region. Sample temperatures were stepped from 263K to 333K in four steps with the sample set to the maximum resistance values at the shuttle tail resistance peak (point-X in Fig. 1b). The resistance was then measured as a function of time without any switching pulses being applied. There was a dramatic exponential decay in time of the resistance at all temperatures, as shown in Fig 3 for the sample at 300K. These results imply that once the forcing function to the out-of-equilibrium state (negative switching pulse) is removed, the system moves by diffusion of the active species (those species that cause the resistivity in the active region to decrease) toward the stable high resistance state.

From the data we postulate that it is oxygen vacancies moving out of the active region and into the bulk by diffusion that are responsible for the time dependent resistance change. Since the oxygen vacancy concentration is reduced near the electrode, the resistance decreases. It should be again noted that the resistance decay is seen only in the 'tail' and the transition regions of the hysteresis loop, and that the high and low states of the sample are stable to $> 10^8$ sec. The decay of resistance found in the shuttle tail (regions HSL-V and HSL-VI in Fig.2a) was consistent with that found for the decay of the resistance in the transition regions (HSL II and HSL-IV in Fig.2a and 2b) of the hysteresis loop implying the same relaxation mechanism active in the transition regions as in the 'tail'.

Therefore, a switching model is proposed for the EPIR effect incorporating oxygen ion/vacancy motion whereby a positive pulse of sufficient magnitude and duration on an electrode will push oxygen ions up against the electrode into the oxygen deficient surface region. These oxygen ions will move into oxygen vacancies thus patching otherwise broken chains of –Mn—O—Mn—O— thereby decreasing the resistance. We have measured the dynamic current density during switching and find that it is very large – $10^1$ to $10^2$ amps/cm$^2$ (Fig. 4), and hence may play a critical role in oxygen/vacancy movement in the interface region of the sample. Note that in Fig 4 current is greatly increased in the transition regions of the hysteresis loop, i.e., where the sample is switching. Enhanced oxygen diffusion by such currents in oxides is a distinct possibility [15]. Especially since PCMO type materials, as strong correlated electron systems, are thought to have phase separation structure at room temperature [9]. This implies lattice instabilities that can easily provide fluctuating smaller activation energies for oxygen motion (paths for oxygen ions).

Likewise a negative pulse of sufficient magnitude and duration will have the opposite effect, moving oxygen vacancies into the interface region thereby breaking the –Mn—O—Mn—O— chains and increasing the resistance up to the original high resistance caused by the oxygen deficient surface of PCMO. Again, the resistance changes occur only in the transition region of the hysteresis loops shown in Fig 1, i.e. in the regions of the loop near the limiting ± maximum voltage pulses.

The complete mechanism of the EPIR effect is probably more complex than only oxygen (vacancy) diffusion due to the strong coupling between electronic, spinic, orbital and elastic degree of freedom in these complex oxide material systems. Increasing oxygen deficiency would increase the $Mn^{3+}/Mn^{4+}$ ratio, the average Mn oxidation state would decrease and thus the average Mn ionic size would increase. As a result the lattice parameter of the oxide may increase in the interface region. Such lattice distortion may influence the electron transport properties of the system. Firstly, a lattice increase would engender structural fluctuations and increase the probability of oxygen diffusion, and thus increase the resistance switching effect. Secondly, the possible presence of inhomogeneous phases with nanometer and/or micrometer scales as discussed in the literature [9,16] may also affect oxygen ions (vacancies) motion in CMR oxides further enhancing the EPIR effect.

To evaluate this oxygen/vacancy diffusion model, we assume the initial concentration of oxygen vacancies in the PCMO at the electrode metal interface before relaxation can be modeled by a half-Gaussian with the maximum value at the interface, and the width given by the thickness of the active region. The solution of the one-dimensional diffusion equation for the oxygen concentration initial condition can be written in closed form [17]:

$$C(x,t) = C(0,0) \cdot L \cdot \frac{Exp[-x^2/(4L^2/4 + D \cdot t)]}{2 \cdot (L^2/4 + D \cdot t)^{1/2}} \quad \text{Eq. 1}$$

Here $C(0,0)$ is the concentration value at the interface, D is the diffusion constant, $D = D_0 \exp(-E/kT)$, L a measure of the width of the initial concentration (we take the thickness of the active layer to be 4L), and t is the time.

We further assume the resistivity in the active region is proportional to the oxygen vacancy concentration plus a constant. To be able to factor out quantities like the $C(0,0)$ and the change in resistivity per oxygen vacancy, we define $R_{max}$ as the value recorded just after the last negative pulse that drives the resistant to its highest value. As noted earlier, we also assume that the 'shuttle tail" part of the hysteresis curve is due to a non-equilibrium concentration of oxygen vacancies driven by the pulse voltage in the active region above the high resistance state, $R_H$.

Using the above assumptions, one can approximate the decay of resistance, R(t) in the 'Shuttle Tail' as a function of time by the following expression:

$$Z(t) = \left(\frac{R_{max} - R_H}{R(t) - R_H}\right)^2 = 1 + 4 \cdot \frac{D}{L^2} \cdot t \quad \text{Eq 2.}$$

Note that Z(t) is defined as a function of experimental parameters only, including the decay of the resistance, R(t), with time. Within the validity of our model, plots of Z(t) vs. time for a given temperature, $T_i$, will yield a straight line (this becomes exact for small values of (4D/L)t) with a slope of $4D(T)/L^2$. A plot of the natural logarithm of the slopes of the straight lines calculated from Equation 2 taken at different absolute temperatures, $T_i$, verses $1/T_i$ will yield the diffusion constant activation energy.

Figure 5a shows curves of Z(t) vs. time for four values of the temperature (around room temperature). Note that while there is considerable noise in the data, the curves approximate straight lines as per Equation 2. Taking the natural logarithm of the slopes of these four Z(t) curves and plotting against 1/T yields Fig 5b. The slope of the best straight line fit results in E/k = 4547.45 K, and an activation energy for diffusion of about 0.4eV.

Since there appears to be no value in the literature of the diffusion constant for oxygen in PCMO single crystals or films at room temperature, we compared our evaluation of the activation energy with oxygen diffusion in other oxides. In particular, the oxygen diffusion constant at room temperature in $Bi_2Sr_2CaCu_2O_{8+d}$ (BSCCO) has been measured as $1.6 \times 10^{-17}$ $cm^2 \cdot sec^{-1}$ by Gramm et al [18]. These authors also determined that the activation energy is about 0.6 eV and $D_0$ is about $2 \times 10^{-7} cm^2 \cdot sec^{-1}$.

The difference in diffusion constant for oxygen that we have found (0.4eV) is significant and indicates that the process responsible for the resistance decay in the transition region and in the Shuttle Peak might be different from oxygen/vacancy diffusion in a lattice under thermal equilibrium. We assume there is a maximum value for local oxygen vacancy concentration that will support the original PCMO lattice. As an estimate, we assume that one could remove up to one oxygen atom from several oxygen octahedra and still maintain the integrity of the unit cell. This assumption yields a maximum local oxygen vacancy concentration that we call $O_{vmax}$. For values less than this, oxygen diffusion in PCMO remains at the very low thermal equilibrium value quoted in the literature for like-structured oxides. Such diffusion yields the time constant controlling long persistency, $t_p \sim 10^8$ sec of the EPIR effect. For oxygen vacancy concentrations greater than $O_{vmax}$, the local lattice becomes highly distorted and strained, leading to a greatly increased diffusion constant. It is this increased oxygen diffusion constant that is measured in the relaxation experiments reported here and controls the time constant $t_r$.

If we assume an active region of about 10-100nm as has been recently observed in our lab by I-AFM techniques [19], we can calculate out the oxygen diffusion constant in that region to be about $2 \times 10^{-15}$ to $5 \times 10^{-14}$ cm$^2$·sec$^{-1}$ ($4D/L^2 = 2 \times 10^{-3}$ sec$^{-1}$ at 300K from fig.5a). Since we have found the activation energy to be 0.4eV, we can calculate $D_0$ to be about $1.1 \times 10^{-8}$ to $2.6 \times 10^{-7}$ cm$^2$·sec$^{-1}$. This prefactor brackets the value found by Gramm [18], indicating similar ionic size and similar attempt frequencies of the diffusing species, i.e. oxygen ions. Further, the change in the diffusion constant is primarily a change in the activation energy, consistent with a distorted lattice that is more open. This oxygen diffusion in a distorted lattice is the relaxation process that occurs in the switching region which includes the 'tail' region. This is reasonable as the switching process is so fast that each pulse leaves the system in a state of temporary non-equilibrium, i.e. even in the non-tail transition region, some octahedra will have more than one missing oxygen ion, and the system simply relaxes back via the more rapid diffusion process reported here.

In summary, we propose that oxygen diffusion is the active agent for resistance switching in PCMO; and that in the interface region the activation energy for oxygen diffusion whose concentration is outside the stable resistance states is about 0.4 eV due to lattice distortion by oxygen vacancies. Within the stable resistance states, oxygen vacancy concentration is less than $O_{vmax}$, so that a larger value of the activation energy, 0.6 eV or higher prevails and makes possible the high persistence time of the non-volatile resistance states. Finally, a simple analysis of the hysteresis curve (Fig 1) during switching, suggests that the mobility for species driven towards (away from) the interface by the application of a switching pulse as calculated by the Einstein equation is greater than diffusion away from (toward) the interface by a factor of $10^6$. This implies that the mobility of oxygen ions (vacancies) associated with the switching pulse is greatly enhanced by current injection during a switching pulse. It is entirely possible that this injection of electrons in the transition region of the hysteresis loop could drastically increase the diffusivity of the oxygen atoms by the required amount. This possibility is consistent with short switching times ($10^{-7}$ sec.) and long persistent times ($3 \times 10^7$ sec – one year). The related phase separation pattern and the stoichiometric change need to be further studied.

Assistance in sample preparation is acknowledged from Mickael Maman. Partial support for this work is acknowledged from NASA, Sharp Laboratories of America, the State of Texas through the Texas Center for Advanced Materials, and the R.A. Welch Foundation.

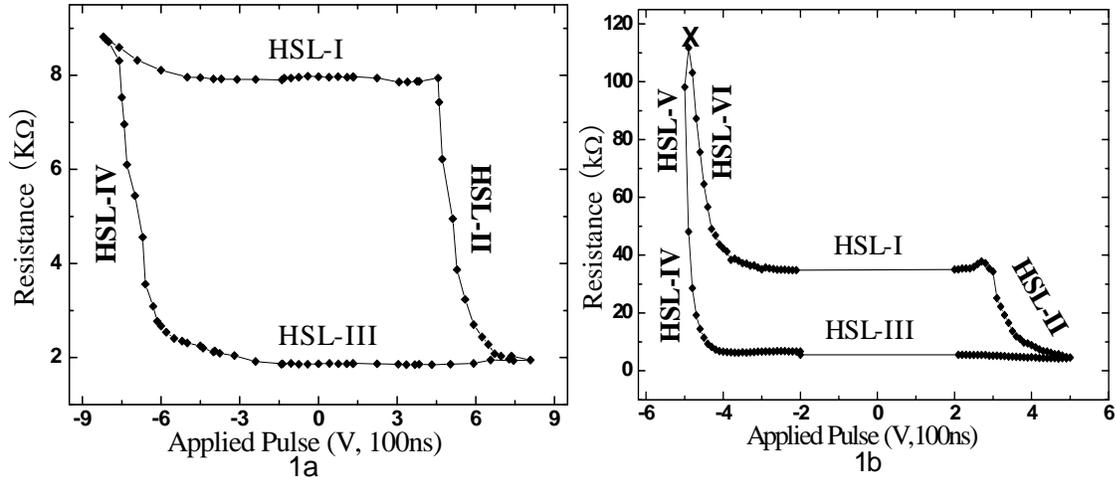

FIG.1 Resistance Hysteresis Switching Loops (HSL) for PCMO EPIR devices: a) hysteresis loop for a PCMO sandwich structure fabricated using a PCMO film grown in an oxygen environment with a YBCO bottom electrode and a Ag top electrode; b) hysteresis loop for a PCMO sandwich structure fabricated using a PCMO film grown in an oxygen-free environment with a Pt bottom electrode and Ag top electrode. (X: the point we start resistance relaxation)

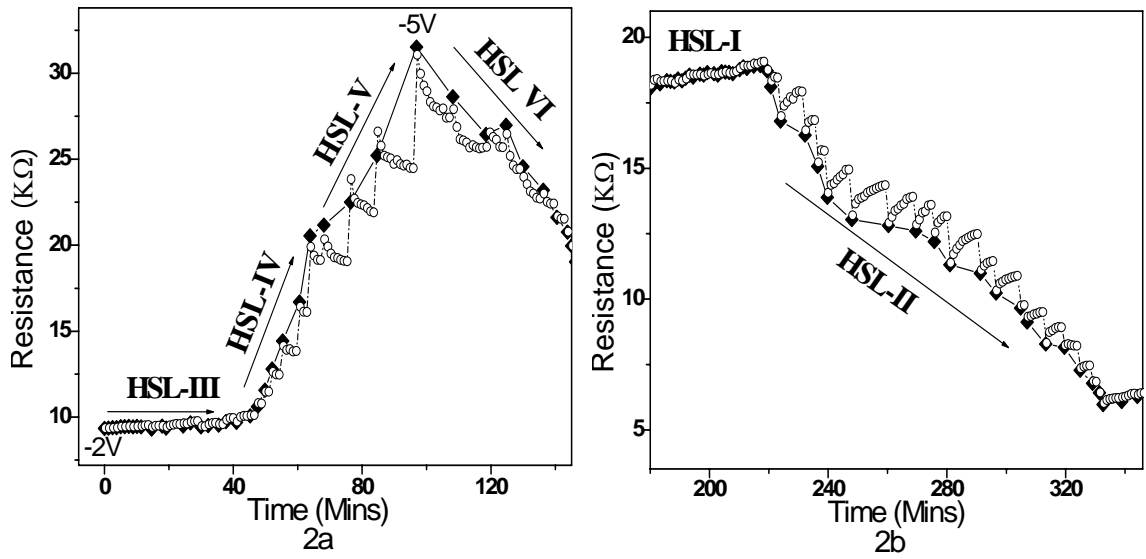

FIG.2 Resistance relaxation (circle) after the application of a pulse (black diamond) for various points in the transition region of the hysteresis switching loop (HSL): a) Negative pulse transition region (HSL-IV/V/VI): resistance going up with R relaxation going down; b) Positive pulse transition region: resistance going down with R relaxation going up (HSL-II).

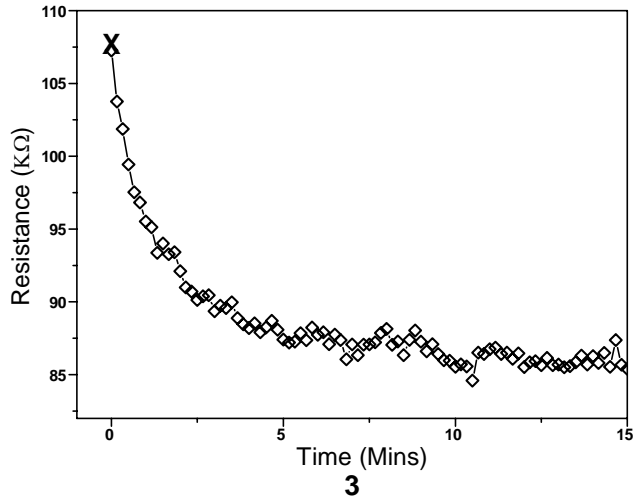

FIG.3 Time decay of the resistance at the peak of the Shuttle Tail ('X' point in Fig.1b) in the hysteresis switching loop (T=297K).

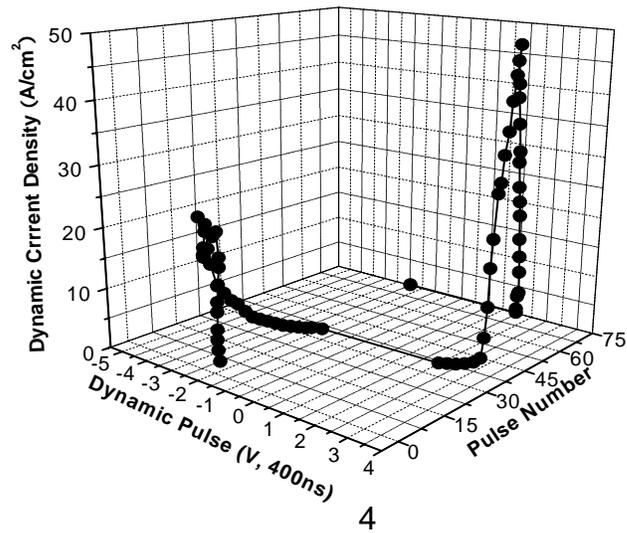

FIG.4 Dynamic current density as a function of pulse voltage around the dynamic hysteresis switching loop.

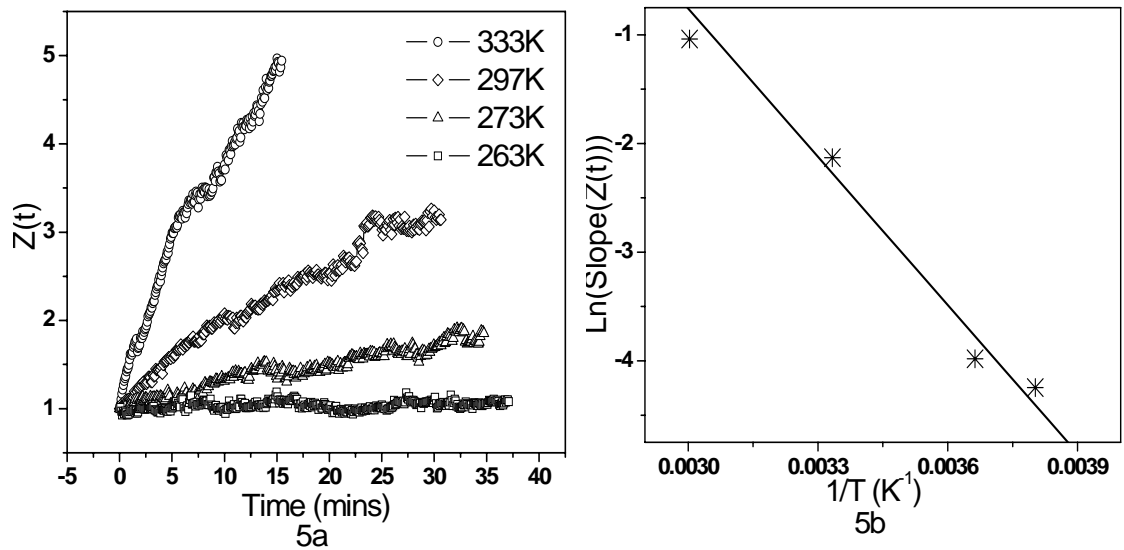

FIG.5  a) Z(t) at different temperatures (333K, 297K, 273K, 263K);  b) Natural log of the slope vs. 1/T with a linear fit (slope: -4547.45K).